30TH INTERNATIONAL COSMIC RAY CONFERENCE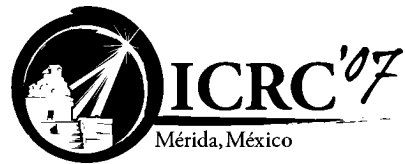

# Elemental Spectra from the CREAM-I Flight

H. S. AHN[1], P. ALLISON[2], M. G. BAGLIESI[3], J. J. BEATTY[2], G. BIGONGIARI[3], P. BOYLE[4],
J. T. CHILDERS[5], N. B. CONKLIN[6], S. COUTU[6], M. A. DUVERNOIS[5], O. GANEL[1], J. H. HAN[7],
J. A. JEON[7], K. C. KIM[1], J. K. LEE[7], M. H. LEE[1], L. LUTZ[1], P. MAESTRO[3], A. MALININE[1],
P. S. MARROCCHESI[3], S. MINNICK[8], S. I. MOGNET[6], S. NAM[7], S. NUTTER[9], I. H. PARK[7],
N. H. PARK[7], E. S. SEO[1,10], R. SINA[1], S. SWORDY[4], S. WAKELY[4], J. WU[1], J. YANG[7], Y. S. YOON[1,10],
R. ZEI[3], S. Y. ZINN[1].
[1] *Inst. for Phys. Sci. and Tech., University of Maryland, College Park, MD 20742 USA*
[2] *Dept. of Physics, Ohio State University, Columbus, Ohio 43210, USA*
[3] *Dept. of Physics, University of Siena and INFN, Via Roma 56, 53100 Siena, Italy*
[4] *Enrico Fermi Institute and Dept. of Physics, University of Chicago, Chicago, IL 60637, USA*
[5] *School of Physics and Astronomy, University of Minnesota, Minneapolis, MN 55455, USA*
[6] *Dept. of Physics, Penn State University, University Park, PA 16802, USA*
[7] *Dept. of Physics, Ewha Womans University, Seoul, 120-750, Republic of Korea*
[8] *Dept. of Physics, Kent State University Tuscarawas, New Philadelphia, OH 44663, USA*
[9] *Dept. of Physics and Geology, Northern Kentucky University, Highland Heights, KY 41099, USA*
[10] *Dept. of Physics, University of Maryland, College Park, MD 20742 USA*
hsahn@umd.edu**Abstract:** The Cosmic Ray Energetics And Mass (CREAM) is a balloon-borne experiment designed to measure the composition and energy spectra of cosmic rays of charge Z = 1 to 26 up to an energy of $\sim 10^{15}$ eV. CREAM had two successful flights on long-duration balloons (LDB) launched from McMurdo Station, Antarctica, in December 2004 and December 2005. CREAM-I achieves a substantial measurement redundancy by employing multiple detector systems, namely a Timing Charge Detector and a Silicon Charge Detector (SCD) for particle identification, and a Transition Radiation Detector and a sampling tungsten/scintillating-fiber ionization calorimeter (CAL) for energy measurement. In this paper, preliminary energy spectra of various elements measured with CAL/SCD during the first 42-day flight are presented.## Introduction

The Cosmic Ray Energetics And Mass (CREAM) balloon-borne experiment is designed to investigate the charge and energy spectra of cosmic-ray nuclei of hydrogen to iron at high energies up to $\sim 10^{15}$ eV. CREAM has had two successful long-duration balloon (LDB) flights, launched from McMurdo Station, Antarctica, for 42 days in 2004-2005 (CREAM-I) and 28 days in 2005-2006 (CREAM-II) [1]. In both flights CREAM employed a 20 radiation length tungsten/scintillating-fiber sampling calorimeter (CAL), preceded by a pair of graphite targets providing $\sim 0.42$ nuclear interaction length, to induce hadronic showers from cosmic-ray nuclei, triggering and measuring the energy of those with energy above $\sim 10^{12}$ eV. Each of the 20 active layers was segmented into 50 one-cm-wide ribbons. Signals from these ribbons were used to reconstruct and extrapolate trajectories back to the Silicon Charge Detector (SCD) of 52 × 56 pixels, for accurate charge measurement. Details of the experiment, including other complementary instruments, namely a Timing Charge Detector and a Transition Radiation Detector, can be found in [2].

Various elements have been studied by analyzing the CREAM-I flight data with CAL/SCD. See [3]

for the spectra of hydrogen and helium. In this paper, preliminary energy spectra of cosmic-ray carbon and oxygen are presented, and compared with results from other experiments.

## Calibration

CAL was placed in one of CERN's SPS accelerator beam-lines, and exposed to a variety of electron, proton, and nuclear fragment beams to verify both the instrument's functionality and the validity of the simulation model. CAL responses to 150 GeV electrons were used for absolute calibration, which is extrapolated to the responses to much higher energy cosmic rays collected during flight [4].

## CREAM-I Flight

During the flight, the payload floated at an average altitude of 128,000 ft, corresponding to a residual atmosphere of 3.9 g/cm$^2$. The analysis in this paper has been performed with only a subset of cosmic-ray events, CAL-triggered by requiring 6 consecutive layers to have energy deposit of more than $\sim 50$ MeV in the highest deposit ribbon, and collected for 23.7 days when both CAL and SCD operation was stable. Live time fraction is assumed to be 75%. The dead CAL channels, noisy SCD pixels, and zero-suppression level in CAL ribbons have been taken into account in the detector simulations.

## Reconstruction

Incident particle trajectory is estimated using $\chi^2$ fitting of a straight line through a combination of CAL hits with highest energy deposit in each layer, in x-z and y-z, respectively. The combination is chosen by rejecting any hit that is not consistent with others to make a straight line. This trajectory is further improved by including, in the fitting, (1) selected CAL hits' neighbors and (2) SCD pixel with highest energy deposit within circle of confusion around the extrapolated position at SCD. This tracking algorithm has been tested with GEANT detector simulations [5]. For the isotropically generated protons within the geometry (particle passing through active SCD area and CAL top/bottom layers, giving geometry factor of 0.37 m$^2$sr) and CAL-triggered, the position resolution at SCD and the tracking efficiency are shown in Fig. 1. The worse position resolution in y-z than in x-z is due to more dead channels.

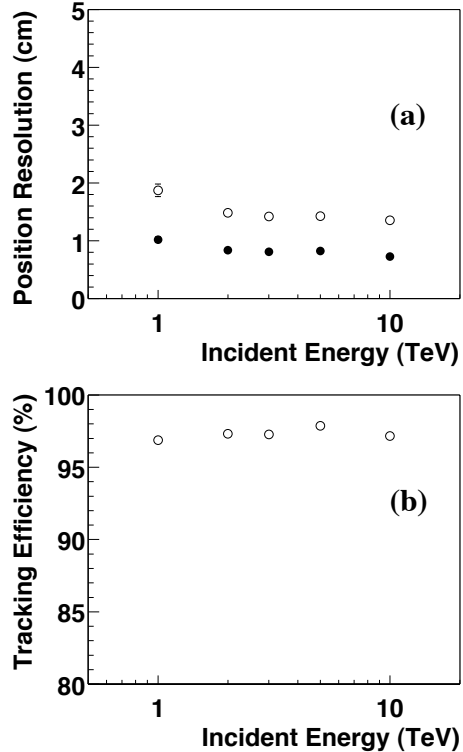

Figure 1: Simulated results of CAL/SCD tracking for protons within the geometry and CAL-triggered, (a) position resolution at SCD, in x-z (filled circles) and y-z (open circles), (b) tracking efficiency.

To determine incident particle charge, the reconstructed trajectory is extrapolated back to SCD, and the highest energy deposit within the circle of confusion is corrected for path-length. The charge is extracted by taking the square root of the corrected signal. In Fig. 2, preliminary charge distribution measured by CREAM-I CAL/SCD shows various elements, including boron, carbon, nitrogen and oxygen. Multiple asymmetric Gaussian functions were applied to parameterize each element, with an exponential function to account for background, the sources of which are still being investigated, including nuclear interactions in the upper detectors and/or support structure before the incident particle reaches the SCD. The contributions from each element and background are also estimated with dashed and dot-dashed lines in Fig. 2.

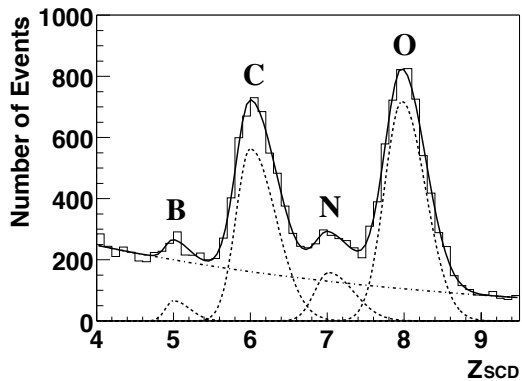

Figure 2: Preliminary charge distribution measured by CREAM-I CAL/SCD for all energies. Overall fitting results (solid line) and the contributions estimated for each element (dashed lines) and background (dot-dashed line) are also shown.

## Absolute Flux Determination

The distribution of energy deposit is divided into 4 bins per decade, and for each bin, the number of carbon/oxygen is estimated by the area of the corresponding asymmetric Gaussian function. To get a distribution of incident energy, detailed study of deconvolution is still in progress, including energy dependence and resolution effects. In this paper, a 0.13% average ratio of energy deposit to incident energy is assumed for energy conversion.

To obtain the differential flux (F) at the top of the atmosphere, the number of incident particles ($N^{\rm inc}$) in each bin of size $\Delta E$, is normalized by

$$F = \frac{N^{\rm inc}}{\Delta E} \times \frac{1}{{\rm GF} \cdot \varepsilon \cdot T \cdot \eta}, \quad (1)$$

where GF is the geometry factor (0.37 m$^2$sr), $\varepsilon$ is a correction for various inefficiencies, T is the live time (17.8 days), and $\eta$ is a correction for atmosphere attenuation. Inefficiencies caused by CAL-trigger, CAL dead channels, and tracking are taken into account in $\varepsilon$. Energy-dependent efficiencies for carbon and oxygen are shown in Fig. 3. $\eta$ is estimated to be 86.3% for carbon, and 85.1% for oxygen, respectively, by calculating the probability of particles within the geometry to survive the atmospheric overburden.

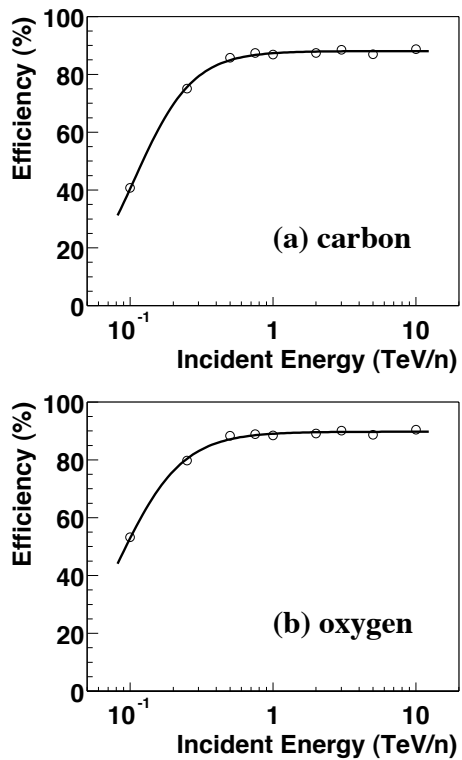

Figure 3: Energy-dependent efficiencies caused by CAL-trigger, CAL dead channels and tracking, for (a) carbon and (b) oxygen.

## Results

Fig. 4 shows the preliminary energy spectra of carbon and oxygen, extracted by CAL/SCD analysis of the CREAM-I flight data (filled circles). Also shown in the plots are the measurements of the HEAO (open stars) [6], the CRN (open crosses) [7], and the CREAM-I Hi-Z (open circles) [8]. De-spite the fact that additional work is underway, including (1) full deconvolution, (2) improved live time estimate, and (3) effect of secondary products from nuclear interaction in the upper detectors and support structure, the results are in fairly good agreement with other observations.

have been presented. The results are in general agreement with other observations and extend the energy range to near 100 TeV/particle. Following this work, analysis on the fluxes of other cosmic nuclei including neon, magnesium, silicon and iron is still in progress.

## Acknowledgements

This work was supported by NASA. The authors thank NASA/WFF, Columbia Scientific Balloon Facility, NSF Office of Polar Programs, and Raytheon Polar Service Company for the successful balloon launch, flight operations, and payload recovery.

## References

[1] E. S. Seo et al., Adv. in Space Res., in press (2007).
[2] H. S. Ahn et al., Nucl. Instrum. Methods A, in press (2007).
[3] Y. S. Yoon et al., OG1.1, this conference.
[4] Y. S. Yoon et al., OG1.5, this conference.
[5] R. Brun et al., GEANT3, CERN DD/EE/84-1 (1984).
[6] J. J. Engelmann et al., Astron. Astrophys. 233, 96 (1990).
[7] D. Müller et al., Astrophys. J. 374, 356 (1991).
[8] S. P. Wakely et al., Adv. in Space Res., in press (2007).

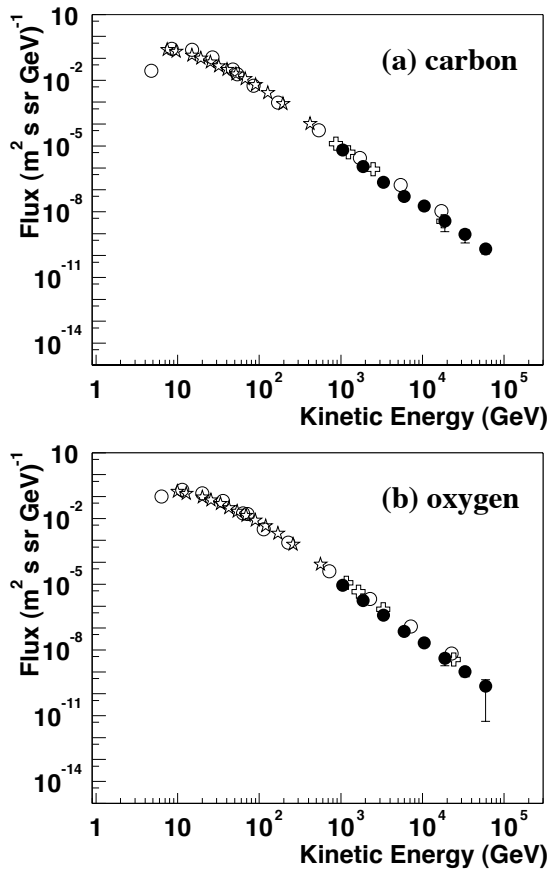

Figure 4: Preliminary energy spectra of (a) carbon and (b) oxygen, measured by CREAM-I CAL/SCD (filled circles), are compared with other measurements. Open stars, crosses, and circles represent the measurement of HEAO, CRN and CREAM-I Hi-Z, respectively.

## Summary

Preliminary energy spectra of carbon and oxygen measured by CREAM-I CAL/SCD, along with the procedure of reconstruction and normalization,